\documentclass{iaus}
\usepackage{graphicx}
\usepackage{rotating}
\usepackage{epstopdf}
\usepackage{epsfig}
\usepackage{natbib}
\usepackage{wasysym}
\usepackage{lscape}
\usepackage{subfigure}

\title[A Visual Guide to Planetary Microlensing] 
{A Visual Guide to Planetary Microlensing}

\author[Leslie A. Rogers \& Paul L. Schechter]   
{Leslie A. Rogers \and Paul L. Schechter}

\affiliation{Department of Physics, Massachusetts Institute of Technology, Cambridge, MA 02139, USA\\email: {\tt larogers@mit.edu, schech@achernar.mit.edu}}
\pubyear{2010}
\volume{276}  
\pagerange{1-2}
\jname{The Astrophysics of Planetary Systems: Formation, Structure, and Dynamical Evolution}
\editors{A. Sozzetti, M. Lattanzi \& A.P. Boss eds.}
\begin{document}

\maketitle

\begin{abstract}
The microlensing technique has found 10 exoplanets to date and promises to discover more in the near future. While planetary transit light curves all show a familiar shape, planetary perturbations to microlensing light curves can manifest a wide variety of morphologies. We present a graphical guide that may be useful when understanding microlensing events showing planetary caustic perturbations.
\end{abstract}

The microlensing approach to discovering planets relies upon chance alignments of two stars along the line-of-sight to Earth (see, e.g., \citet[]{Bennett08} or \citet[]{Gaudi10} for  a detailed review). During the near-alignments the foreground star acts as a gravitational lens, bending light from the background source star. A point mass foreground lens star produces two images of the source: a positive parity image outside the Einstein radius, and a negative parity image inside the Einstein radius. Together, the two unresolved images result in an overall magnification, $A_0$, of the source. As the source and lens stars move relative to one another, the overall magnification and image positions vary with time. 
If the lens star harbors a planet that lies near the path of one of the images, the planet can perturb the magnification, ($A = A_0+\delta A$), producing an observable signature in the microlensing light curve.

\begin{figure}[b]
\begin{center}
\includegraphics[width=0.89\textwidth]{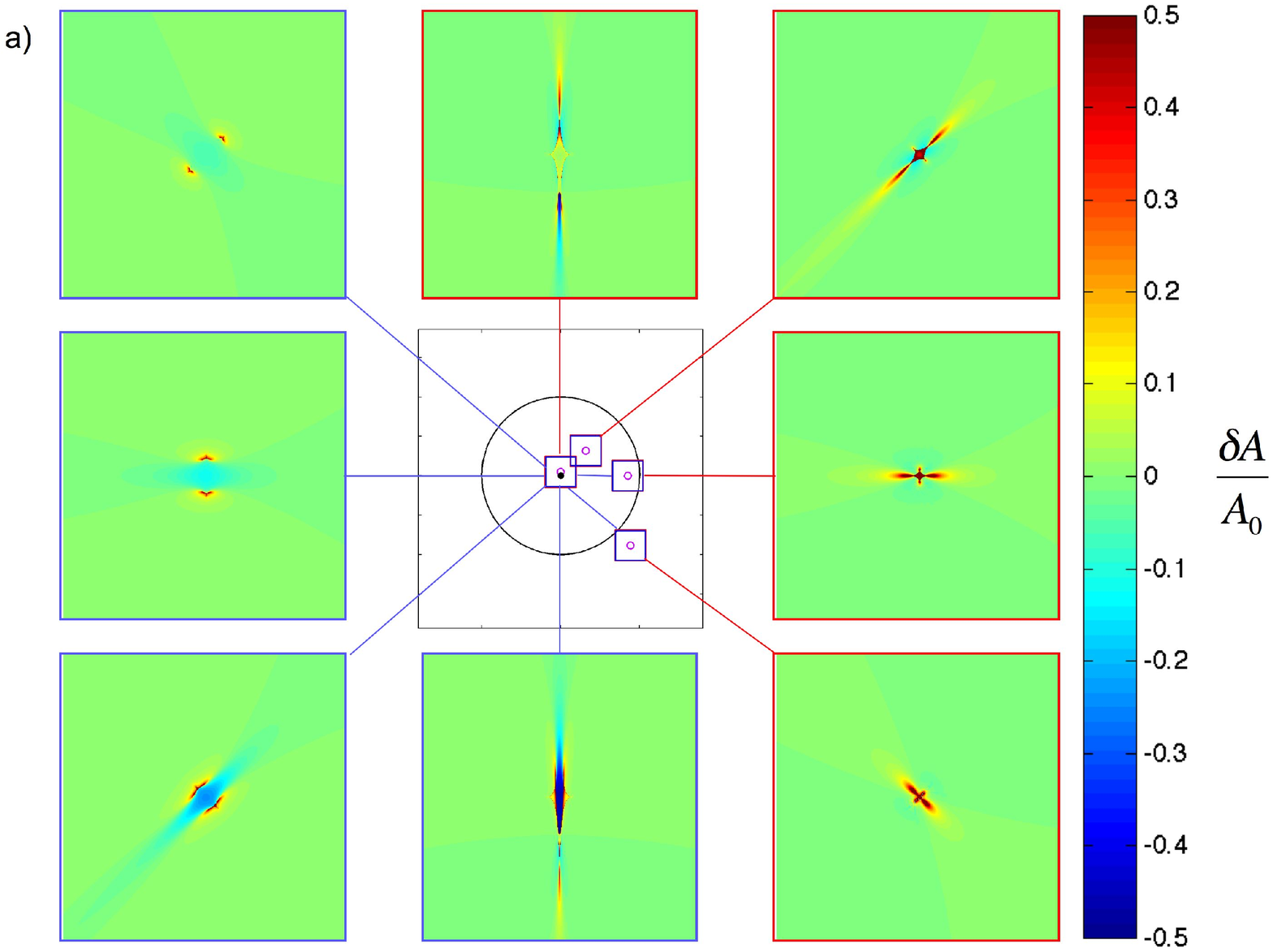} 
\includegraphics[width=0.89\textwidth]{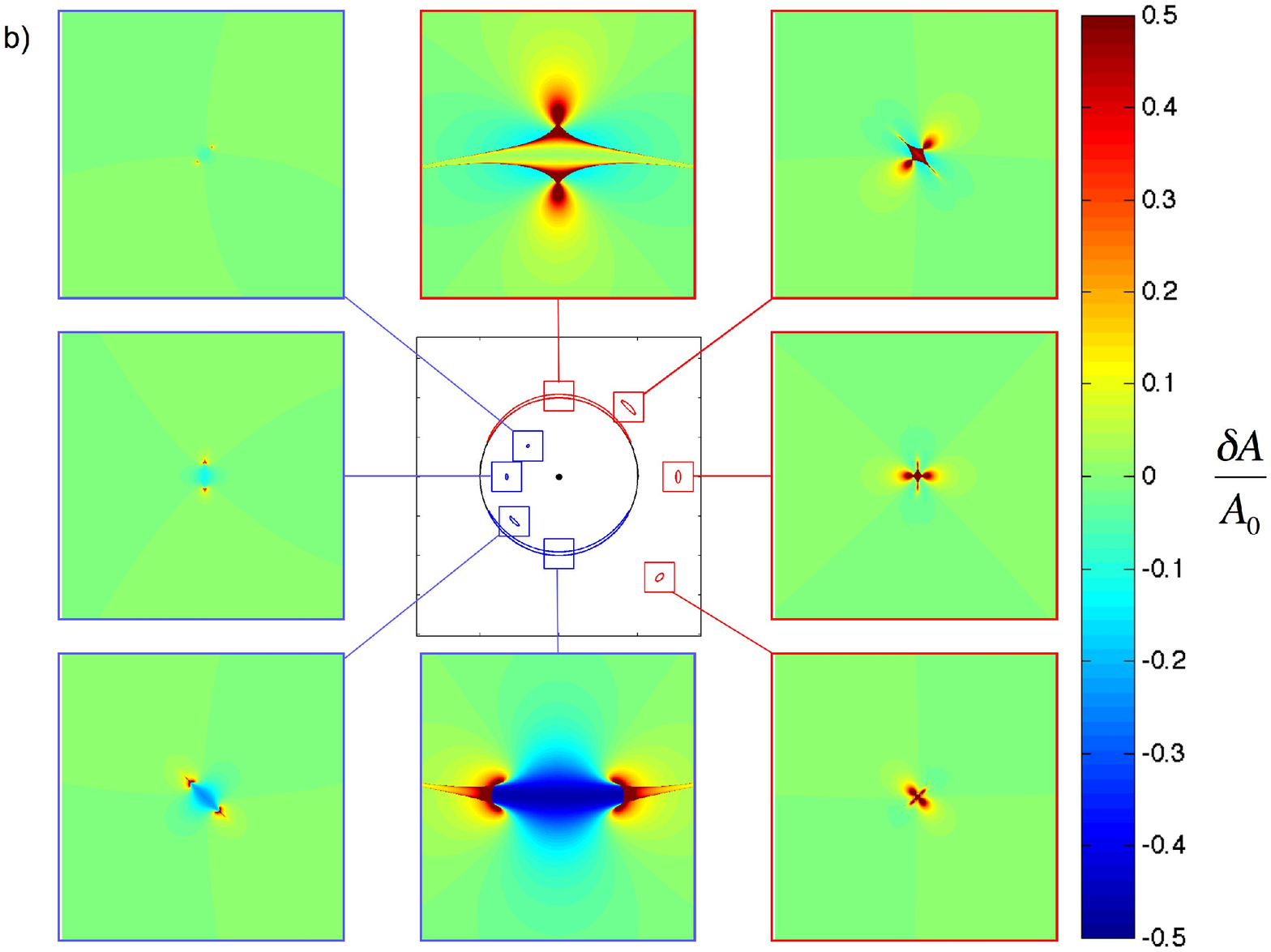} 
 \caption{ Fractional change in magnification of a point source induced by a planet around the lens star ($M_p=10^{-4} M_*$) relative to the point mass lens (no planet) case. In the magnification map color scale red and yellow denote magnification increases, shades of blue denote magnification decreases, and green denotes magnifications relatively unchanged by the presence of the planet. In a) we plot how the magnification varies with the position of the source star for a fixed planet position, while in b) we show how the magnification depends on the planet position in the lens plane for fixed source position.}
   \label{fig:1}
\end{center}
\end{figure}

We present  a compact yet comprehensive graphic that illustrates the range of possible magnification maps for microlensing events showing planetary caustic perturbations (Figure~\ref{fig:1}). The diagrams in the middle panels illustrate the lens configurations considered in the surrounding maps. The black circle represents the primary lens star's Einstein ring, while the squares outline the regions spanned by the magnification maps. The center of each magnification map corresponds to a Òdirect-hitÓ, for which the planet falls exactly on one of the unperturbed images of the source lensed by the point-mass primary. Magnification maps on opposite sides of the middle panels correspond to the same Òdirect-hitÓ source position (0.05, 0.45, 0.85 or 1.25 Einstein radii from the lens star); the planet perturbs the positive parity image in the red-outlined maps, and the negative parity image in the blue-outlined maps. 

The source-lens-planet configurations at magnification map center are identical between Figures~\ref{fig:1} a) and b). In a) the source position varies across the map with the planet fixed, while in b) the planet position varies across the map with the source fixed. Tracing the source trajectory through the source plane maps in a) yields the microlensing light curve. In contrast, the maps in b) reveal the region of the lens plane within which one is sensitive to planets at a given instant in time.   


\end{document}